\pdfoutput=1

\documentclass[article,10pt,onecolumn,superscriptaddress,bibnotes,showpacs,showkeys,longbibliography]{revtex4}
\usepackage{microtype}
\usepackage{graphicx}
\usepackage{amsmath}
\usepackage{amssymb}
\usepackage[colorlinks,linkcolor=blue,urlcolor=blue,citecolor=blue]{hyperref}

\usepackage{mathptmx}
\usepackage{times}
\usepackage{epsfig,amsopn}
\usepackage{graphicx}
\usepackage{bm,amsmath,amssymb}
\usepackage{natbib}
\usepackage{braket}
\usepackage{float}
\usepackage{enumerate}
\usepackage{hyperref}
\allowdisplaybreaks[1]
\newcommand{\D}{{\rm d}}

\begin{document}
\def\be{\begin{equation}}
\def\ee{\end{equation}}
\def\bea{\begin{eqnarray}}
\def\eea{\end{eqnarray}}
\def\f{\frac}
\def\l{\label}
\def\nn{\nonumber}

\definecolor{dgreen}{rgb}{0,0.7,0}
\def\redw#1{{\color{red} #1}}
\def\green#1{{\color{dgreen} #1}}
\def\blue#1{{\color{blue} #1}}
\def\brown#1{{\color{brown} #1}}

\newcommand{\eref}[1]{~(\ref{#1})}%
\newcommand{\Eref}[1]{Equation~(\ref{#1})}%
\newcommand{\fref}[1]{Fig.~\ref{#1}} %
\newcommand{\Fref}[1]{Figure~\ref{#1}}%
\newcommand{\sref}[1]{Sec.~\ref{#1}}%
\newcommand{\Sref}[1]{Section~\ref{#1}}%
\newcommand{\aref}[1]{Appendix~\ref{#1}}%
\newcommand{\sgn}[1]{\mathrm{sgn}({#1})}%
\newcommand{\erfc}{\mathrm{erfc}}%
\newcommand{\Erf}{\mathrm{erf}}%

\title{ Diffusion under time-dependent resetting}

\author{Arnab Pal}
\affiliation{Schulich Faculty of Chemistry, Technion-Israel Institute of Technology, Haifa 32000, Israel} 

\author{Anupam Kundu}
\affiliation{International center for theoretical sciences, TIFR, Bangalore 560012, India}

\author{Martin R. Evans}
\affiliation{ SUPA, School of Physics and Astronomy, University of Edinburgh, Mayfield
Road, Edinburgh EH9 3JZ, United Kingdom}



\begin{abstract}
{We study a Brownian particle
diffusing under  a time-modulated stochastic resetting mechanism to a
fixed position.
The rate of resetting $r(t)$ is a function of the time $t$ since the
last reset event.
We derive a sufficient condition on $r(t)$ for a steady-state 
probability distribution of the position of the particle to exist.
We derive the form of the steady-state distributions  under some particular choices of $r(t)$
and  also consider the late time relaxation behavior of the
probability distribution.
We consider first passage time properties for the Brownian
particle to reach the origin
and derive a formula for the mean first passage time.
Finally, we study optimal properties of the mean first passage time and show
that a threshold function is at least locally optimal for the problem
of minimizing the mean first passage time.}
\end{abstract}


\maketitle


\section{Introduction}
\label{sec:Introduction}
Over the last few decades there has been great interest in the study of search problems that appear 
in various contexts from animal foraging, protein binding on DNA,
internet search algorithms to locate one's misplaced keys.
In such different situations, one important issue is to consider optimal search strategies. 
For example, in protein binding on DNA, a protein molecule binds to a 
specific target binding site on DNA by an appropriate mixture of 3D
diffusion and 1D sliding motion on DNA -- called facilitated diffusion \cite{Sokolov05}.

Quite often efficient search strategies involve a mixture of local steps and long-range moves. Such search strategies are 
called intermittent search process and are observed 
in foraging animals such as humming birds or bumblebees \cite{Bell-91,Stanleybook}. The E. Coli bacteria 
alternatively uses ballistic moves, called ``runs'' with random
``tumbles'',  to change its direction in order to reach high food
concentration regions \cite{berg08}.

Recently, an intermittent stochastic strategy containing such a mixture
of local and long-range moves has been introduced \cite{MZ99,EM11a,
  EM11b} in which diffusion of a particle is interrupted by
stochastically resetting it to a preferred position whereupon the diffusion
process starts afresh.  Examples of such stochastic resetting are
found in a wide variety of situations. In daily life, while searching
for some lost possession, after an unsuccessful search for some duration
of time one often goes back to the starting place and recommences the
search process again.  In the ecological context, the movement of
animals during foraging period often involves local diffusive search
for food \cite{Turchin, Stanleybook}.  Interestingly, such local
diffusive movements are interrupted by long range moves to relocate
themselves in other areas and after that local diffusive motion
restarts from the relocation region\cite{Benichou-11}. Movement of
free-ranging capuchin monkeys in the wild is described quite well by
random walks with preferential relocations to places visited in the
past \cite{Boyer}.  There are other examples in nature where similar
notions of stochastic resetting can also be found.  For example, in
the biological context several living organisms use stochastic switching
between different phenotypic states to adapt in a fluctuating
environment \cite{Kussel05, Kusselgene, Reingruber, Visco-10,
  Reingruber11}.  Stochastic restarts are often considered as a useful 
 strategy to optimize computer search algorithms in hard
combinatorial problems \cite{Lovasz, Montanari, Konstas}. Also in the context of population growth, random 
catastrophic events may cause a sudden reduction in the population size and reset it to some lower value.
\cite{Visco-10}.

The mechanism of stochastic resetting
fundamentally affects the properties of diffusion process.  Consider a
particle, starting from $x_0$, diffusing in one dimension. Let $x(t)$
be its position at time $t$. Along with diffusion, the particle is
subject to a resetting mechanism in which its motion is interrupted
stochastically such that the position of the particle is reset to some
fixed position $x_r$ at some rate $r$ and after every such event the
particle recommences its diffusive motion.  In the absence of resetting,
the Gaussian distribution of the position of the particle never
reaches a steady state. The width of the distribution keeps on growing
with time as $\sim \sqrt{t}$.  On the other hand, in the presence of
resetting the distribution of the position becomes a globally
current-carrying non-equilibrium steady state with non-Gaussian
fluctuations. Resetting also has another important consequence on the
properties of diffusion process : the mean first passage time (MFPT)
to a particular position of a diffusing particle is infinite whereas
it becomes finite in presence of resetting. In fact there is an
optimal resetting rate $r$ for which this finite MFPT is minimum.  

In the last decades, several theoretical works have been dedicated to the study of first-passage properties of
intermittent search processes, and in particular to the question of the minimization of the mean first-passage
time i.e. the mean time to locate some target \cite{First-Passage-Book,BCMSV05,TVB12,CBV15,BGV15,GM15}. 
These works use different analytical and numerical approaches to minimize the MFPT in a broad range of 
contexts  such as search processes mixing slow diffusive movement and 
fast ballistic motion \cite{BCMSV05}, persistent random walks \cite{TVB12}, 
finding targets in a domain \cite{CBV15} and Brownian search in spatial heterogeneous media \cite{GM15}.
Also in \cite{KMSS14}, an intermittent search process with Levy flights
interrupted by random resettings has been studied where a first order
phase transition associated to a discontinuous change in the optimal
parameters has been observed.

Recently a number of generalizations of the simple diffusion with
resetting have been made.  The generalization from the
one-dimensional case to higher dimensions has been considered in
\cite{EM14}. In the context of a target search process, the effect of partial absorption
has been considered in \cite{Whitehouse:2013}. 
Properties of non-equilibrium steady state for 
diffusion with resetting have been studied in the presence of a potential 
\cite{Pal14} or in a bounded domain \cite{Chistou15}. Other generalizations 
include resetting to the current maximum of the Brownian particle \cite{MSS15-2}, 
resetting in continuous-time random walks \cite{Montero-13}, in L\'evy flights \cite{KMSS14} etc. 
The effect of resetting in the dynamics of interacting multi-particle systems such as fluctuating interfaces \cite{GMS14}, 
coagulation-diffusion process \cite{Henkel:2014} 
and in general chemical reaction schemes \cite{RRU15} have been studied.

In this paper we consider a different generalization of the resetting
process:
we consider a resetting rate  $r(t)$ that  is {\em time-dependent}.
This generalization is quite natural in the context
of target search. While searching for a lost object, the searcher
naturally would not like to reset in the beginning of the search, but
as time progresses without success, the searcher would be more and more
inclined to go back to the most likely location and start the search
process anew. As a result it would be quite natural to consider a
situation where the resetting rate $r(t)$ grows from zero as time increases. 
Specifically, we consider a situation where the resetting rate  depends on the {\em time
since the last reset}. Thus the rate as well as the searcher
is reset at the resetting event.

In the case of time-independent resetting rates, the \textit{renewal property} of Brownian motion has been exploited 
to compute the propagator (the probability density of the particle
being at $x$ at time $t$ given that it began at $x_0$ at time 0) as well as the steady-state distribution of the position of the particle 
\cite{EM11a,EM11b,EM14,SabhapanditTouchette15}. 
More precisely, the propagator has been expressed as an integral over the time of the \textit{last resetting events}. 
For time dependent resetting rates, we find that it is more convenient to express the propagator of the particle 
as an integral over the time of \textit{first resetting events}. We refer to this formalism as the {\it first renewal picture} 
in contrast to  the formalism used in \cite{EM11a,EM11b,EM14,SabhapanditTouchette15} which we refer to as the {\it last renewal picture}. We first recapitulate the constant 
resetting rate case in Sec. \ref{recap-const-r} where we re-derive the known results using the {\it first renewal picture}. 
Then in Sec. \ref{time-dep-r} we discuss the generalization of this approach to the time dependent case, 
where we compute the Laplace transform (w.r.t. time) of the propagator for a general resetting rate function $r(t)$.

Next we study the steady-state properties of the Brownian particle
subject to such a resetting mechanism.  In the absence of resetting,
the Brownian particle keeps on diffusing freely over space with time;
but the introduction of a resetting mechanism creates a current
towards the reset position. This current may balance the outward
current (flowing towards infinity) in the large time limit. As a
result the system may reach a steady state eventually but {\it not}
for all possible choices of the rate function $r(t)$. The question is
then for what choices of $r(t)$ will a steady state be attained?  We
address this question in Sec. \ref{SS-time-dep-r}, where we study the
steady-state behavior of the Brownian particle in presence of a
time-dependent resetting rate. We find a sufficient condition on the
late time behavior of $r(t)$ for a steady state to be attained which is
given by equation (\ref{sscond}). Such a condition can be easily
understood as follows : Let us consider the following two extreme
cases, a) $r(t) \to \infty$ and b) $r(t)=0$. In the former case, the
particle is always instantaneously reset to the resetting position
$x_r$ and hence the probability distribution of the position is then
given by $\delta(x-x_r)$ irrespective of the initial position $x_0$.
On the other hand, when $r(t)=0,~\forall t$, the particle diffuses
freely and never attains a steady state distribution (on an infinite
system). Hence if $r(t)$ decays fast enough to zero as time increases,
then the particle may have finite probability of not being reset at
all and as a result the distribution of the position of the particle
may not reach a steady state. In fact we find that if $r(t)$ decays
slower than $\sim 1/t$ for large $t$, then the particle reaches a
steady state. The next question is: how does it relax to the steady
state, if indeed one exists?

This question was first studied by Majumdar \emph{et al} \cite{MSS15}
in the context of constant resetting rate 
case. They found that for a given large time $t$, 
the distribution in an inner core region around the reset position has already relaxed to the (time-independent) steady state whereas the distribution 
in the outer region has not yet relaxed. 
The front dividing the relaxed from the non-relaxed region moves linearly with time through the system. In Sec. \ref{sec:ltd} 
we observe the same phenomena for a specific class of time dependent resetting rate function $r(t) \sim t^{\theta},~\theta>-1$. 
However, depending on the value of $\theta$, the motion of the front becomes linear, superlinear or sublinear with $t$.

One of the main motivations for studying Brownian motion with
resetting is to improve our understanding of  search paradigms,
as discussed in the first
and second paragraphs of the introduction. As a quantitative measure
of the performance of a search process, one may consider the mean search time or the mean
first passage time (MFPT) $T(x_0)$ to a static target 
for a given starting position $x_0$ and diffusion constant $D$. 
In the absence of resetting, the MFPT of a
free Brownian particle to a static target at origin is infinite. On
the other hand, in the presence of resetting to the initial position with
constant rate $r(t)=r_c$, the MFPT becomes finite \cite{EM11a, EM11b}.
Moreover, it has also been shown \cite{EM11a, EM11b} that there exists
an optimal choice of the constant rate $r^*_c$ for which MFPT $T(x_0)$
becomes minimum. In this work we ask the following questions : can
this scenario be improved further if one considers a time-dependent
resetting rate $r(t)$ ?  What is the optimal rate function $r(t)$ for
which the MFPT $T(x_0)$ becomes minimum ? We address these questions
in Sec. \ref{survival_probability} and Sec. \ref{optimal-r(t)}
respectively. We show that indeed there exists a few choices of
$r(t)$ for which it is possible to achieve a  MFPT lower than the minimum
MFPT obtained using constant resetting rate $r^*_c$. In
Sec. \ref{optimal-r(t)} we study the optimal resetting rate function,
where we conjecture that the optimal time-dependent rate function is given by
a threshold function.

We note that several very recent works
have also considered the problem of time-dependent resetting.
Eule and Metzger \cite{Eule15}
use a generalized (non-Markovian) master equation approach
to consider the  stationary distribution  of the particle position
and also some late time properties. In that work
 a Gamma-distribution of waiting times between successive reset events 
is considered and mean first passage times 
are numerically determined.
Nagar and Gupta \cite{NG15} consider the particular case of  
a power law distribution  of waiting times between resets.
They  discuss the steady states  and some first passage time properties.
In contrast, in  our work we consider the general scenario of a time-dependent resetting rate
$r(t)$
(with corresponding waiting time distribution $r(t) {\rm e}^{-R(t)}$
given by (\ref{rate1})) for various different choices of $r(t)$. Furthermore we consider the late time  relaxation of the probability distribution
and we study analytically the problem of optimising the mean first passage time.

\section{The model}
\noindent
First let us define diffusion with resetting with time-dependent
resetting rate. We consider a single particle (or searcher)
in one dimension with initial position $x_0$ at $t=0$ and a resetting  position $ x_r$.
The position $x(t)$ of the particle at time
$t$ is updated by the following stochastic
rule: in a small time interval $t \to t+ \D t$ 
the position  $x(t)$ becomes 
\begin{eqnarray}
 x(t+dt) & =& x_0 \quad {\rm with\,\,probability}\,\, r(t-\tau_l)\, dt 
 \\
& =&  x(t) + \eta(t) dt \quad {\rm with\,\,probability}\,\,
     (1-r(t-\tau_l)\, dt) \label{rule.1}
\end{eqnarray}
where $\tau_l$ is the time of the last resetting event. Thus the
resetting rate $r(t-\tau_l)$ is a function of the {\em time elapsed since
the last resetting event}.
In (\ref{rule.1})  $\eta(t)$ is a Gaussian white noise with mean 
and  two-time correlator given by
\be
\langle\eta(t)\rangle=0,
~~\langle\eta(t)\eta(t')\rangle=2D\delta(t-t'),
\ee
where $D$ is the diffusion constant.
Here, angular brackets denote averaging over noise realizations.
The initial condition is $x(0) =x_0$.
To simplify matters, from now on we shall take the initial position to coincide
with the resetting position
\be
x(0)=x_{0}  =x_r\;,
\ee
unless otherwise specified. The dynamics thus  consists of a stochastic mixture of resetting to
the initial position with rate $r(t-\tau_l)$ (long range move) and ordinary diffusion 
(local  move) with diffusion constant $D$.

We define  
$P_r(x,t|x_0,0)$ as the probability of finding the particle at position $x$
at time $t$, given that it was at $x_0$ at time $t=0$ 
in the presence of \textit{time-independent} stochastic resetting,
$\mathbb{P}_r(x,t|x_0,0)$ being the same  quantity for the \textit{time-dependent} stochastic resetting.
In both cases the suffix `$r$'  indicates the presence of resetting.


\subsection{Recap of constant resetting rate $r(t)=r_c$}
\label{recap-const-r}
\noindent
For completeness we review here the formalism and the results for the case of
constant resetting rate $r(t)=r_c$ \cite{EM11a,EM14}.
In this case, one can write down a Master equation for $P_r(x,t|x_0,0)$
from the dynamical rules for the evolution of the particle given in the
preceding section \cite{EM11a} 
\bea
\f{\partial P_r}{\partial
t}=D\f{\partial^{2}P_r}{\partial
x^{2}}-r_cP_r+r_c\delta(x-x_0),
\l{transient}
\eea
with the initial condition $P_r(x,t=0|x_0,0)=\delta(x-x_{0})$. For convenience, 
we have omitted the arguments of $P_r(x,t|x_0,0)$ in the above equation. 
Here, the second and third terms on the right hand side (RHS) account for the resetting
events, denoting, respectively,  the negative probability flux $-r_cP_r$ from each point $x$ and a
corresponding positive probability flux into $x=x_r=x_0$. The steady-state solution for the time-independent case $P_r^{\rm st}(x)$ satisfies
\bea
0=D\f{d^{2}P_r^{\rm st}}{d x^{2}} -r_cP_r^{\rm st}+r_c\delta(x-x_0).
\l{SS1}
\eea

Alternatively, a renewal picture which we refer to as the {\em last
 renewal picture}  may be used to write down an equation for
$P_r(x,t|x_0,0)$ in terms of the free propagator  $G(x,t|x_0,0)$
for a pure diffusive process (without resetting) as in \cite{EM11a,EM14}
\bea
P_r(x,t|x_0,0)&=&e^{-r_ct}G(x,t|x_0,0) \nn \\
&+&r_c\int_{0}^{t}d\tau_l~e^{-r_c(t-\tau_l)}~G(x,t-\tau_l|x_0,0).
\l{LR1ti}
\eea
Here, we have divided the  process into two contributions. 
The first term in the RHS signifies that there has been no reset at all between time $(0,t)$
with the probability of no resets given
by   ${\rm e}^{-r_ct}$. This probability  is then simply  multiplied
by the free Brownian propagator
\bea
G(x,t|x_0,0)=\f{1}{\sqrt{4\pi Dt}}~\exp[-\f{(x-x_0)^2}{4Dt}]\;.
\label{BP1}
\eea
The second term in the RHS accounts for the fact that
there can be multiple resets: the integral sums over contributions
in which the \textit{last resetting} event
takes place between time $\tau_l$ and $\tau_l+d\tau_l$.
Subsequently, the particle propagates freely until the observation
time scale $t$. It can be shown that \eref{LR1ti}
satisfies the master equation \eref{transient}.
The steady state can be obtained by taking the
infinite time limit
\bea
P^{\rm st}_r(x)=r_c\int_{0}^{\infty}d\tau_l~e^{-r_c \tau_l}~G(x,\tau_l|x_0,0)~,
\label{rtist}
\eea
which satisfies the steady state equation
given by \eref{SS1}. The integral (\ref{rtist}) can be evaluated to
yield \cite{EM14}  
\begin{equation}
P^{\text{st}}_r(x)
= ~\frac{\alpha_0}{2}~{\rm e}^{-\alpha_0|x-x_0|}~,~~\text{where},~~\alpha_0=\sqrt{\f{r_c}{D}}~.
\l{alpha_0}
\end{equation}
The last resetting equation (\ref{LR1ti}) also allows the long time
relaxation to the steady state to be analyzed \cite{MSS15} as we shall
review in Section~\ref{sec:ltd}.

In this work we introduce a \textit{first renewal picture}
where instead of the last resetting, we consider the \textit{first
resetting} between time $\tau_f$ and $\tau_f+d\tau_f$ having started
from $t=0$.
Subsequently, the particle diffuses from $\tau_f$ until time $t$,  in the
presence of multiple resets. It is again
 straightforward to write down an equation for the probability
\bea
P_r(x,t|x_0,0)&=&e^{-r_ct}G(x,t|x_0,0) \nn \\
&+&r_c\int_{0}^{t}d\tau_f~e^{-r_c(t-\tau_f)}~P_r(x,t|x_0,\tau_f)~,
\l{LR2}
\eea
where the first term in the RHS corresponds to 
 trajectories in which there are no resets at all.  The integral in the
second term sums over trajectories in which  there
has been a first reset between time $\tau_f$ and $\tau_f+d\tau_f$
and then there can be multiple resets which is taken care of 
by the reset propagator $P_r(x,t|x_0,0)$ inside the integral. 
The equivalence between \eref{LR1ti} and
\eref{LR2} is easy to show
by taking  Laplace transforms
of both equations.  We find that they 
result in identical expression in the Laplace space
\bea
\tilde{P}(x,s|x_0,0)=\frac{r_c+s}{s}~\tilde{G}(x,r_c+s|x_0,0)~,
\label{EQV1}
\eea
where $\tilde{P}_r(x,s|x_0,0)$ is the Laplace transform of $P_r(x,t|x_0,0)$
and $\tilde{G}(x,s|x_0,0)$ is the Laplace transform of $G(x,t|x_0,0)$. Taking 
the inverse Laplace transform with respect to $s$, one can easily obtain ~(\ref{LR1ti}).


\subsection{Time-dependent resetting rate}
\label{time-dep-r}
\noindent
We now turn to the main subject of this paper, that of
time-dependent resetting rates $r(t)$, as defined above.
In this case  one cannot simply write a Master equation for $\mathbb{P}_r(x,t|x_0,0)$ in the
presence of time-dependent resetting rate.
This is because one must in addition keep track of the time since the last reset.
The renewal pictures  are  more useful than the Master equation formalism
to describe the time-dependent rate process.
To this end, we define the following
time-integrated quantity 
\bea
R(\tau)=\int_{0}^{\tau}d\tau' r(\tau').
\l{rate1}
\eea
Then the probability of no resets subsequent to an initial reset at
$t=0$ is given by ${\rm e}^{-R(t)}$
and $r(t){\rm e}^{-R(t)}$ is the probability density for a first reset
to occur in the interval $t \to t+ dt$.

In the case of time-dependent resetting the \textit{last renewal} equation (\ref{LR1ti}) is  modified to
\bea
\mathbb{P}_r(x,t|x_0,0)&=&e^{-R(t)}G(x,t|x_0,0) \nn \\
&+&\int_{0}^{t}d\tau_l~\psi(\tau_l) e^{-R(t-\tau_l)}~G(x,t-\tau_l|x_0,0)
\l{LR1}
\eea
where  $\psi(\tau_l)$ is the probability density for a reset (which
turns out to be the last)  to occur
in $t \to t+ dt$. We shall return to this picture in section~\ref{sec:ltd} 
where we consider the late time relaxation behavior.

However, to study the steady-state behavior it is most convenient to
use the \textit{first renewal} framework.
In the presence of the time-dependent 
reset process the \textit{first renewal} equation becomes
\bea
\mathbb{P}_r(x,t|x_0,0)&=&e^{-R(t)}~G(x,t|x_0,0)  \nonumber \\
&+&\int_{0}^{t}d\tau~r(\tau)~e^{-R(\tau)}~\mathbb{P}_r(x,t-\tau|x_0,0)\;.
\l{propagator1}
\eea
By taking the Laplace transform, we obtain
\bea
\tilde{\mathbb{P}}_r(x,s|x_0,0)=\f{\tilde{\mathbb{Q}}(x,s|x_0,0)}{s\tilde{\mathbb{H}}_{r}(s)}\;,
\l{lpropagatorlt1}
\eea
where we have defined
\bea
\tilde{\mathbb{P}}_r(x,s|x_0,0)&=&\int_{0}^{\infty}dt~e^{-st}\mathbb{P}_r(x,t|x_0,0) ,\label{Plt} \\
\tilde{\mathbb{Q}}(x,s|x_0,0)&=&\int_{0}^{\infty}dt~e^{-st}~e^{-R(t)}G(x,t|x_0,0) ,\label{Q_mbb_s}\\
\tilde{\mathbb{H}}_{r}(s)&=&\int_{0}^{\infty}dt~e^{-st}~e^{-R(t)}.
\l{lpropagator2}
\eea
Equation (\ref{lpropagatorlt1})  is the main result of this section. In principle, the Laplace
transform can be inverted although in practice this is difficult for arbitrary $r(t)$.


\section{Steady state behaviour}
\label{SS-time-dep-r}
\noindent
In the previous section we have studied the propagator of a
  Brownian particle in presence of a  time-dependent resetting mechanism 
using the {\it first renewal picture}. We have computed the Laplace transform (\ref{lpropagatorlt1}) of the
propagator $\mathbb{P}_r(x,t|x_0,0)$ in terms of the Laplace transforms of the free propagator $G(x,t|x_0,0)$ weighted by the probability $e^{-R(t)}$ of 
no reset in duration $t$. 
In this section we are interested in the large time behavior of
$\mathbb{P}_r(x,t|x_0,0)$. More precisely, we are interested in
whether   
$\mathbb{P}_r(x,t|x_0,0)$ takes a time independent form in $t \to \infty$ limit. If so then, in the Laplace transform language, 
this simply means that $\tilde{\mathbb{P}}_r(x,s|x_0,0)$ should be
expressible as $1/s$ multiplied by a quantity that does not depend on $s$ in 
$s \to 0$ limit and the steady state distribution would be given by 
\bea
\mathbb{P}_r^{\text{st}}(x)= \lim_{s \to 0}~\f{\tilde{\mathbb{Q}}(x,s|x_0,0)}{\tilde{\mathbb{H}}_{r}(s)}~,
\l{stst}
\eea
provided the limit exists and is not zero.
Now it is easy to check from the definitions  (\ref{Q_mbb_s}) and (\ref{lpropagator2})
that if $\tilde{\mathbb{H}}_{r}(s \to 0) < \infty$  
then also $\tilde{Q}(x,s
\to 0|x_0,0) < \infty$.
Thus a    sufficient condition
on the choice of $r(t)$ to achieve a steady state is that
$\tilde{\mathbb{H}}_{r}(s \to 0) < \infty$  
which implies that
\begin{equation}
\int_0^\infty {\rm e}^{-R(t)} dt < \infty\;.
\label{sscond}
\end{equation}
This condition implies that $R(t)$ must grow sufficiently fast for
large $t$ so that  ${\rm e}^{-R(t) }\to 0$ sufficiently quickly.

Recalling the definition 
$R(t)=\int_{0}^{t}d\tau r(\tau)$
(\ref{rate1}),  condition (\ref{sscond}) will hold if $r(t)$ is an
increasing function of time since then $R(t) \to \infty$.
Thus for an increasing resetting rate, there
always exists a steady state. Also if $r(t)$ tends to some constant
 value greater than zero for large $t$, then $R(t)$ grows linearly with time and
again (\ref{sscond}) is satisfied.
On the other hand if $r(t)$ is a  function decreasing to zero at large  time since
the last reset, to achieve a steady state, it must
decrease sufficiently slowly that $R(t)$ diverges with time.
Thus (\ref{sscond}) implies that
if the decreasing rate function $r(t)$
is bounded one fold i.e. if
$r(t)$ decays \textit{more slowly} than $\f{1}{t}$, then there exists a 
unique steady state.

In the following, we explore few plausible choices of the 
time-dependent rates.


\subsubsection{Case I: linear rate }
\noindent
As mentioned in the introduction, it is natural in the context of search
processes to consider a rate  which is an increasing function of
time. The simplest such function is a linear function.
For the linear resetting process $r(t)=b_0 t$,
we find the following relations
\begin{align}
R(t)~=&~\f{b_0~t^2}{2}~,\nonumber \\
\tilde{\mathbb{H}}_{r}(s)~=&~\sqrt{\f{\pi}{2b_0}}~e^{s^2/2b_0}~\erfc[\f{s}{\sqrt{2 b_0}}],\nonumber \\
\tilde{\mathbb{Q}}(x,s|x_0,0)~=&~\int_{0}^{\infty}~dt~e^{-st}~e^{-b_0 t^2/2}\f{1}{\sqrt{4\pi Dt}}~\exp[-\f{(x-x_0)^2}{4Dt}]~.
\label{details-h}
\end{align}
One can then write the steady state,
in terms of a simple integral  depending on one variable
by inserting  the relations (\ref{details-h}) in \eref{stst}
\begin{eqnarray}
\mathbb{P}_r^{\text{st}}(x)~&=&~\frac{\sqrt[4]{b_0}}{\sqrt{4\sqrt{2}\pi D }
}~\left(\frac{d}{d \ell}J(\ell)\right)_{\ell=\frac{|x-x_0|\sqrt[4]{b_0}}{\sqrt{4\sqrt{2}\pi D }}}~, \label{P_st-lin} \\
 \text{where},~~J(\ell)&=&\int_0^\infty dz~e^{-z^2}\text{erf}\left(\frac{\ell}{\sqrt{z}} \right). \label{J(l)}
\end{eqnarray}
This integral can be computed using {\it Mathematica} and its 
full expression is given in terms of hypergeometric functions in Appendix \ref{sec:app}.
In \fref{figr1}a, we compare this result against the same obtained from 
 direct numerical simulation of the dynamics and observe an excellent agreement.

\begin{figure*}
\includegraphics[scale=0.62]{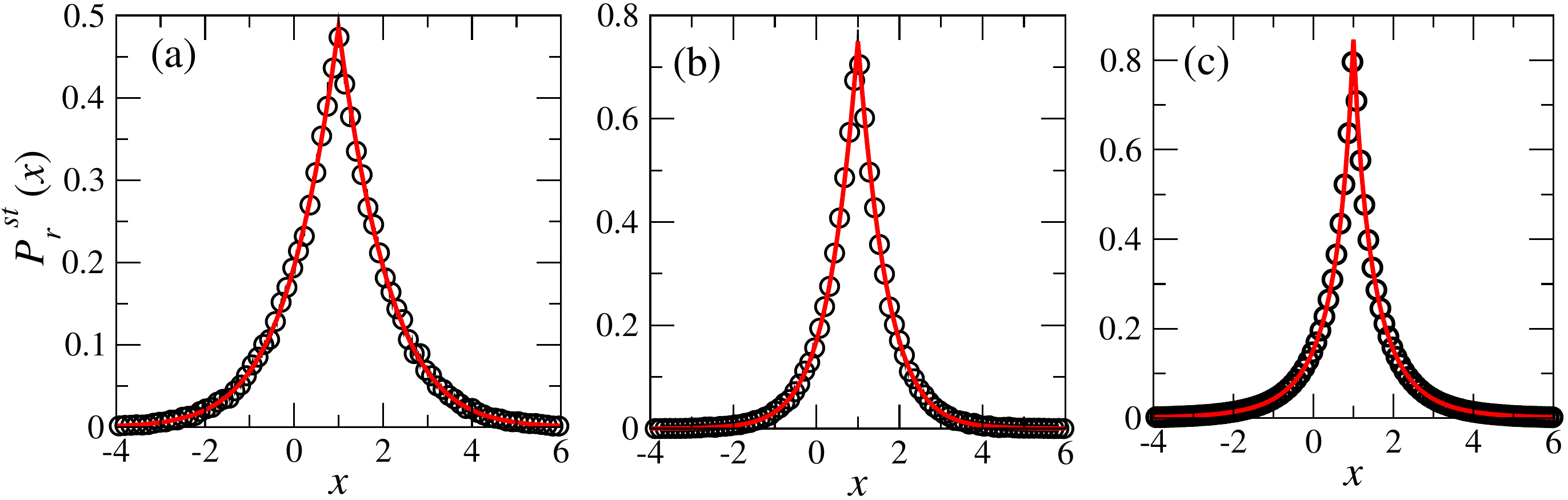}
\caption{(Color online) Stationary distributions $\mathbb{P}_r^{\text{st}}(x)$
of a single Brownian particle moving on a  one-dimensional line in the
presence of  time-dependent resetting. 
In this plot we consider three choices of the resetting rate functions $r(t)$: (a) linearly increasing $r(t)=b_0~t$ and 
(b) increasing but bounded  $r(t)=b_0(\f{1}{\epsilon}-\f{1}{t+\epsilon})$ and (c) decreasing rate $r(t)=\frac{b_0}{\sqrt{t}}$, where $b_0$ is some parameter in 
appropriate units. In all the figures the circles represent data obtained 
from direct numerical simulations whereas the solid lines correspond to the respective theoretical expressions (\ref{P_st-lin}), (\ref{P_st-ibb}) and (\ref{P-ss-th-lt-1})
in the main text.
The intrinsic parameters of the system are : $D = 1$, $x_r=x_0=1$ and $b_0 = 1~$( in respective units ) for all the three cases.}
\l{figr1}
\end{figure*}

\subsubsection{Case II:  increasing, but bounded rate }
\noindent
One unrealistic feature of an increasing rate, such as the linear rate
considered in the previous paragraph, is that it may increase
indefinitely. Here we consider 
a resetting processes in which the rate is an increasing function
but  bounded by an upper limit:
\bea
r(t)=b_0\left(\f{1}{\epsilon}-\f{1}{t+\epsilon}\right)\;.
\eea
Thus $r(0)=0$ and $r(t) \to b_0/\epsilon$ as $t\to \infty$.
In this case, the integrated rate $R(t)$  (\ref{rate1}) is given by $R(t)~=~b_0 ({t}/{\epsilon})- b_0 \log[1+{t}/{\epsilon}]$ implying 
$e^{-R(t)} \to 0$ for $t \to \infty$. Hence we expect to have a  steady state which can be computed 
from ~(\ref{stst}). The function
$\tilde{\mathbb{H}}_{r}(s)$ can be easily computed and its 
limit as $s\to 0$ is given by 
\begin{equation}
 \tilde{\mathbb{H}}_{r}(s)|_{s\to 0}~=~\epsilon e^{b_0-(1+b_0)\ln b_0} \Gamma_g(1+b_0,b_0)~,
\end{equation}
where $\Gamma_g(u,v)=\int_v^\infty dw~w^{u-1}e^{-w}$ is an incomplete Gamma-function.
To complete the evaluation of $\mathbb{P}_r^{\text{st}}(x)$, we now need to 
compute $\tilde{\mathbb{Q}}(x,s|x_0,0)$ from \eref{Q_mbb_s} in the $s \to 0$ limit.
After some straightforward manipulations one 
can show that $\tilde{\mathbb{Q}}(x,s|x_0,0)|_{s \to 0}$ is given by 
\begin{eqnarray}
&&\tilde{\mathbb{Q}}(x,s|x_0,0)|_{s \to 0} = \frac{\sqrt{\epsilon}}{4\sqrt{D}}~\left( \frac{d}{d \ell} \mathcal{J}_{b_0}(\ell)\right)_{\frac{|x-x_0|}{\sqrt{4D\epsilon}}},~~\text{with}
\nonumber \\
&&\mathcal{J}_{b_0}(\ell)= \int_0^\infty dz~e^{-b_0z}(1+z)^{b_0}\text{erf}\left(\f{b_0}{\sqrt{z}} \right), \label{J_b_0(l)} \\
&&= \sum_{k=0}^\infty \sum_{m=1}^\infty \f{(-1)^{k}}{k!m!}(-2\ell)^m(b_0)_k\left(\f{m-2}{2}\right)_k b_0^{m/2-k-1}.
\end{eqnarray}
where the symbol $(a)_k$ is given by $(a)_k=a(a-1)...(a-k+1)$. Hence, the steady state distribution $\mathbb{P}_r^{\text{st}}(x)$ is given 
by 
\begin{equation}
 \mathbb{P}_r^{\text{st}}(x) = \frac{e^{(1+b_0)\ln b_0-b_0}}{ \Gamma_g(1+b_0,b_0)}~\frac{1}{4\sqrt{D\epsilon}}~
 \left( \frac{d}{d \ell} \mathcal{J}_{b_0}(\ell)\right)_{\frac{|x-x_0|}{\sqrt{4D\epsilon}}}. \label{P_st-ibb}
\end{equation}
In \fref{figr1}b, we  compare the analytical results against the same obtained from 
 direct numerical simulation of the dynamics. The results are found to be
in excellent agreement with the numerical results.


\subsubsection{Case III:  Rates decreasing as a  power law  $r(t)\sim b_0/t^{\theta}$}
\noindent
We now explore the case  where the rate $r(t)$ decreases as a function of time. If
the rate decreases too quickly  then we expect from (\ref{sscond}) that a steady state
will not be attained. Therefore it is of natural interest to inquire more
details of the  condition for which the steady states exist and consequently their
characteristic forms if they do.

We first consider
the case  $\theta<1$ for which we may take
\begin{equation}
r(t) = b_0/t^{\theta}\quad\forall t
\end{equation}
(it turns out that the singularity at $t=0$ does not affect the steady
state).
Then we find
\begin{equation}
R(t) =  b_0 t^{1-\theta}/(1-\theta) \quad\mbox{and}\quad R(t) \to \infty \quad \mbox{as}\quad
t \to \infty\;.
\end{equation}
In  this case, criterion (\ref{sscond}) is satisfied and 
we have steady states  given by 
\begin{eqnarray}
\mathbb{P}_r^{\text{st}}(x)&=& \frac{\tilde{\mathbb{Q}}(x,0|x_0,0)}{\tilde{\mathbb{H}}_{r}(0)}
\label{P-ss-th-lt-1}
\end{eqnarray}
where,
\begin{eqnarray}
  \tilde{\mathbb{H}}_{r}(0)~&=& b_0^{-1/(1-\theta)}
                                (1-\theta)^{\theta/(1-\theta)}
                                \Gamma\left(\frac{1}{1-\theta}\right), \nonumber \\
\tilde{\mathbb{Q}}(x,0|x_0,0)&=&~
\int_{0}^{\infty}~\D t~e^{-b_0 t^{1-\theta}/(1-\theta)}\f{\exp[-\f{(x-x_0)^2}{4Dt}]}{\sqrt{4\pi Dt}}~. 
\end{eqnarray}
We now consider the case for which $\theta>1$. In order to ensure
convergence of $R(t)$ we take
\begin{equation}
r(t) =b_0/(t+\epsilon)^{\theta}\label{pleps}\;.
\end{equation}
The time-integrated rates can be found as
\begin{align}
R(t)~\simeq&~\frac{b_0~t^{1-\theta}}{1-\theta}~+ \mbox{constant},\nonumber \\
e^{-R(t)}~&\to \mbox{constant} > 0 ~~\text{as}~{t\to\infty}
\end{align}
and therefore criterion (\ref{sscond}) is not satisfied. Moreover, it can be checked that
for small $s$, 
  $\tilde{\mathbb{H}}_{r}(s) \sim 1/s$
whereas  $\tilde{\mathbb{Q}}(x,s|x_0,0)\sim s^{-1/2}$ so that 
(\ref{stst}) $\to 0$ as $s\to 0$ and there is no steady state.

Finally, we notice that, $\theta=1$ is a marginal
case.
Taking $r(t)$ as in (\ref{pleps})
\begin{equation}
r(t) =b_0/(t+\epsilon)
\end{equation}
 we obtain
\begin{equation}
e^{-R(t)}~=\left[ \frac{\epsilon}{t+\epsilon}\right]^{b_0}\;.
\end{equation}
For $b_0 \leq 1$,
  $\tilde{\mathbb{H}}_{r}(s)$ diverges as $s \to 0$
whereas  $\tilde{\mathbb{Q}}(x,s|x_0,0)$   converges as $s\to 0$. Thus
according to (\ref{stst}) there is no steady state.
However, if  $b_0>1$,
 $\tilde{\mathbb{H}}_{r}(s)$ converges  as $s \to 0$, thus condition
(\ref{sscond}) is satisfied and there is a steady state given
 by (\ref{stst}) with
\begin{eqnarray}
  \tilde{\mathbb{H}}_{r}(0)~&=& \frac{\epsilon}{b_0-1}, \nonumber \\
\tilde{\mathbb{Q}}(x,0|x_0,0)&=&~
\int_{0}^{\infty}~dt~\left[\frac{\epsilon}{t+\epsilon}\right]^{b_0}\f{\exp[-\f{(x-x_0)^2}{4Dt}]}{\sqrt{4\pi Dt}}~. 
\end{eqnarray}
In \fref{figr1}c, we plot the steady state distribution corresponding to the
power law choice of $r(t) = b_0/t^{\theta}$ with $\theta=1/2$ and compare
with the same obtained from numerical simulations. We again observe an excellent agreement.


\section{Late time relaxation of the probability distribution}
\label{sec:ltd}
\noindent
For the case of time-independent constant resetting rate $r_c$, the late time
relaxation of the probability distribution has been studied in
\cite{MSS15}.  It was shown that at large time $t$ an inner region of
the distribution $|x-x_0| < (4Dr_c)^{1/2}t$ has relaxed to the
(time-independent) non-equilibrium steady state, whereas the outer region
$|x-x_0| > (4Dr_c)^{1/2}t$ has not yet relaxed and the time-dependent
probability is dominated by trajectories that have not yet undergone any
resetting. Thus a front dividing the `equilibrated' region from the `unequilibrated'
one moves linearly with time through the system with speed
$v=(4Dr_c)^{1/2}$\cite{MSS15}.

Here we extend this analysis to the case of time-dependent resetting
rates.
We begin with the  last renewal equation (\ref{LR1})
\bea
\mathbb{P}_r(x,t|x_0,0)&=&e^{-R(t)}G(x,t|x_0,0) \nn \\
&+&\int_{0}^{t}d\tau_l~\psi(\tau_l) e^{-R(t-\tau_l)}G(x,t-\tau_l|x_0,0),
\l{LR1new}
\eea
and assume that the
second term in the right hand side (RHS) dominates at large $t$
\begin{eqnarray}
\mathbb{P}_r(x,t|x_0,0)&\simeq &
\int_{0}^{t}d\tau_l~\psi(\tau_l) e^{-R(t-\tau_l)}~G(x,t-\tau_l|x_0,0) \nonumber\\
&=& 
t \int_{0}^{1}dw~\psi((1-w)t) e^{-R(wt)}~G(x, wt|x_0,0) \label{ltP}\nonumber \\
&= &\left(\frac{t}{4\pi D}\right)^{1/2}
\int_{0}^{1}\f{dw}{\sqrt{w}}~\psi((1-w)t) {\rm e}^{-R(wt)   -
     \frac{(x-x_0)^2}{4Dwt}} 
\end{eqnarray}
where in the second line, a new integration variable  $w$ is defined through $\tau_l = (1-w)t$.
We now wish to evaluate the integral by the saddle-point method for
large $t$.

Consider a class of rate processes such as the following
\begin{equation}
r(t) = b_0 t^\theta\quad\mbox{where}\quad \theta > -1 ~.
\l{rt-form}
\end{equation}
This ensures us of both increasing and decreasing resetting rates along
with the existence of a steady state in each case.
Now, we have the integrated quantity
\begin{equation}
R(wt)  = \frac{b_0 w^{1+\theta}}{1+\theta} t^{1+\theta}\;.
\end{equation}
We observe in the scale where $|x-x_0|= u t^{1+\theta/2}$
i.e. we consider  $|x-x_0| \to \infty$ and $ t\to \infty$ with $u$ held fixed.
 The integral
 (\ref{ltP}) becomes
\begin{eqnarray}
\mathbb{P}_r(x,t|x_0,0)&\simeq &
\left(\frac{t}{4\pi D}\right)^{1/2}
\int_{0}^{1}  \f{dw}{\sqrt{w}} \psi((1-w)t) {\rm e}^{-t^{1+\theta}\left[
\frac{b_0 w^{1+\theta}}{1+\theta}
   + \frac{u^2}{4Dw}\right]}\;. \nonumber \\
\l{ltP-2}
\end{eqnarray}
Now we expect $\psi \to constant < \infty$ as $t\to \infty$
and therefore
we may evaluate the integral in \eref{ltP-2} by the saddle-point method. 
The saddle-point equation simply reads
\begin{eqnarray}
\frac{\partial }{\partial w}  \left[
\frac{b_0 w^{1+\theta}}{1+\theta}
   + \frac{u^2}{4Dw}\right]
= b_0 w^\theta -\frac{u^2}{4Dw^2} =~0~,
\end{eqnarray}
which defines the value $w^*$ of $w$ that dominates the integral
\begin{equation}
w^* =  \left( \frac{u^2}{4 D b_0} \right)^{1/(2+\theta)}\;.
\end{equation}
However this value can only dominate the integral  if it is within the integration
range $0\leq w^* \leq 1$ which is the case if $u^2 <  4 Db_0$ or equivalently
\begin{equation}
|x-x_0| < \sqrt{4 D b_0}\, t^{1+\theta/2}\;.
\label{front}
\end{equation}
On the other hand, if  $w^*$ is out of the integration range  in which case (\ref{front}) no
longer holds and it turns out that the
first term in the right hand side (RHS) of the last renewal equation (\ref{LR1new}) dominates. This
contribution essentially represents those trajectories of the dynamics
which  have not participated in the resetting process yet.


\begin{figure*}
\includegraphics[width=.29\hsize]{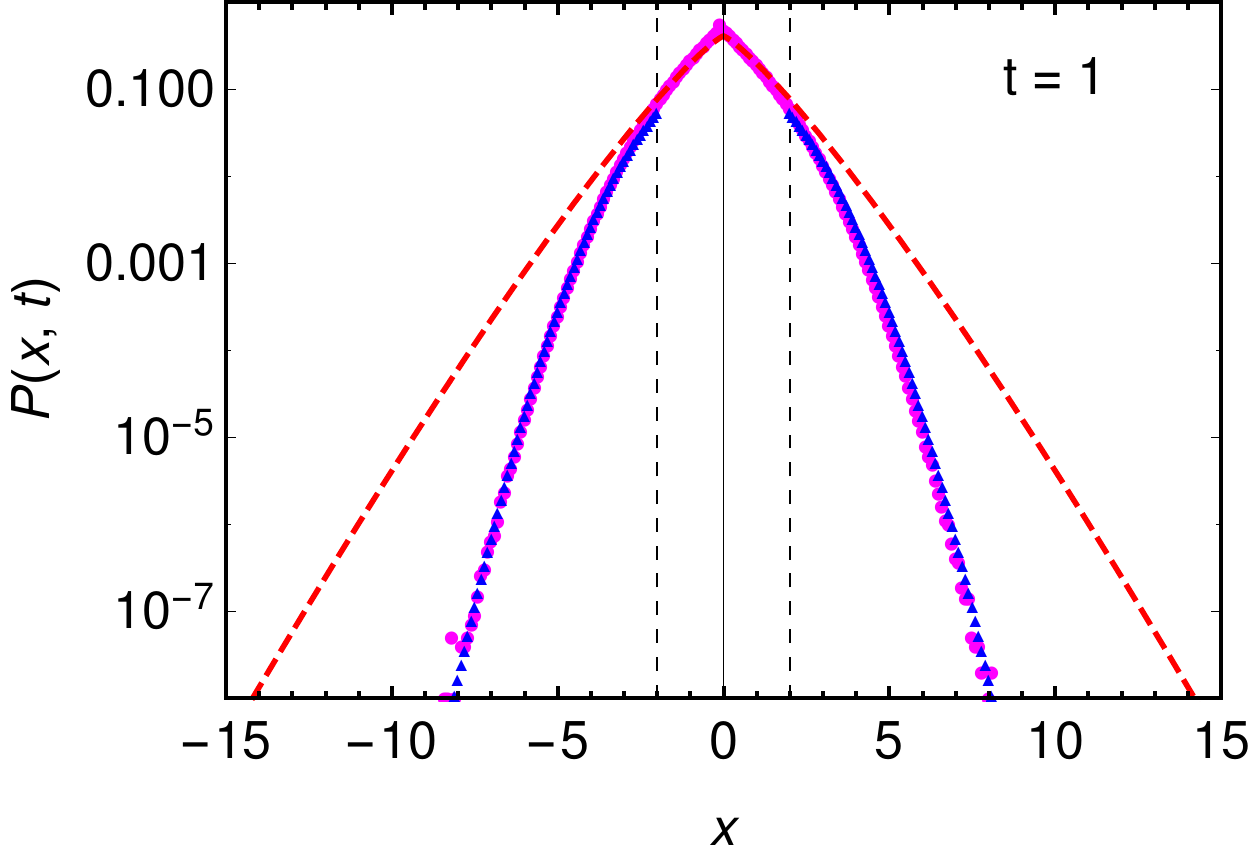}
\includegraphics[width=.285\hsize]{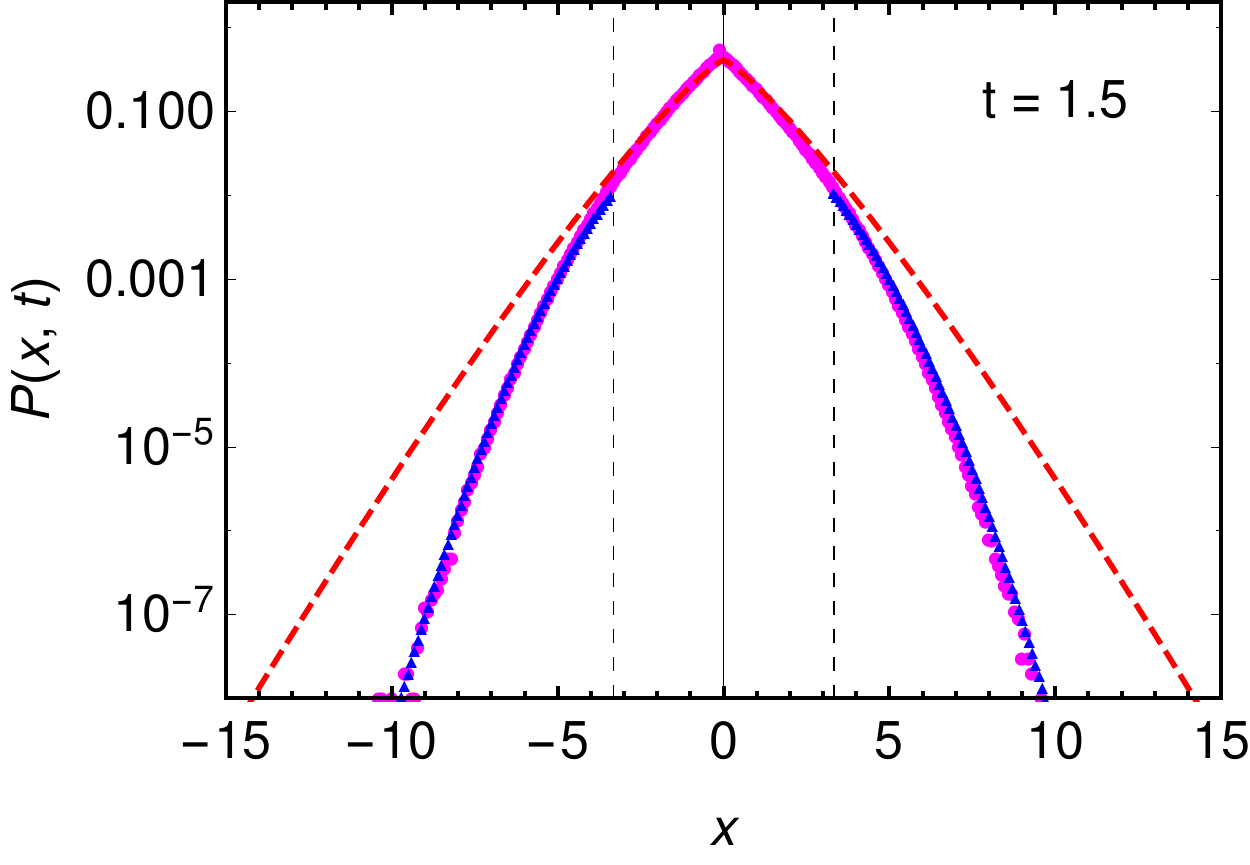}
\includegraphics[width=.285\hsize]{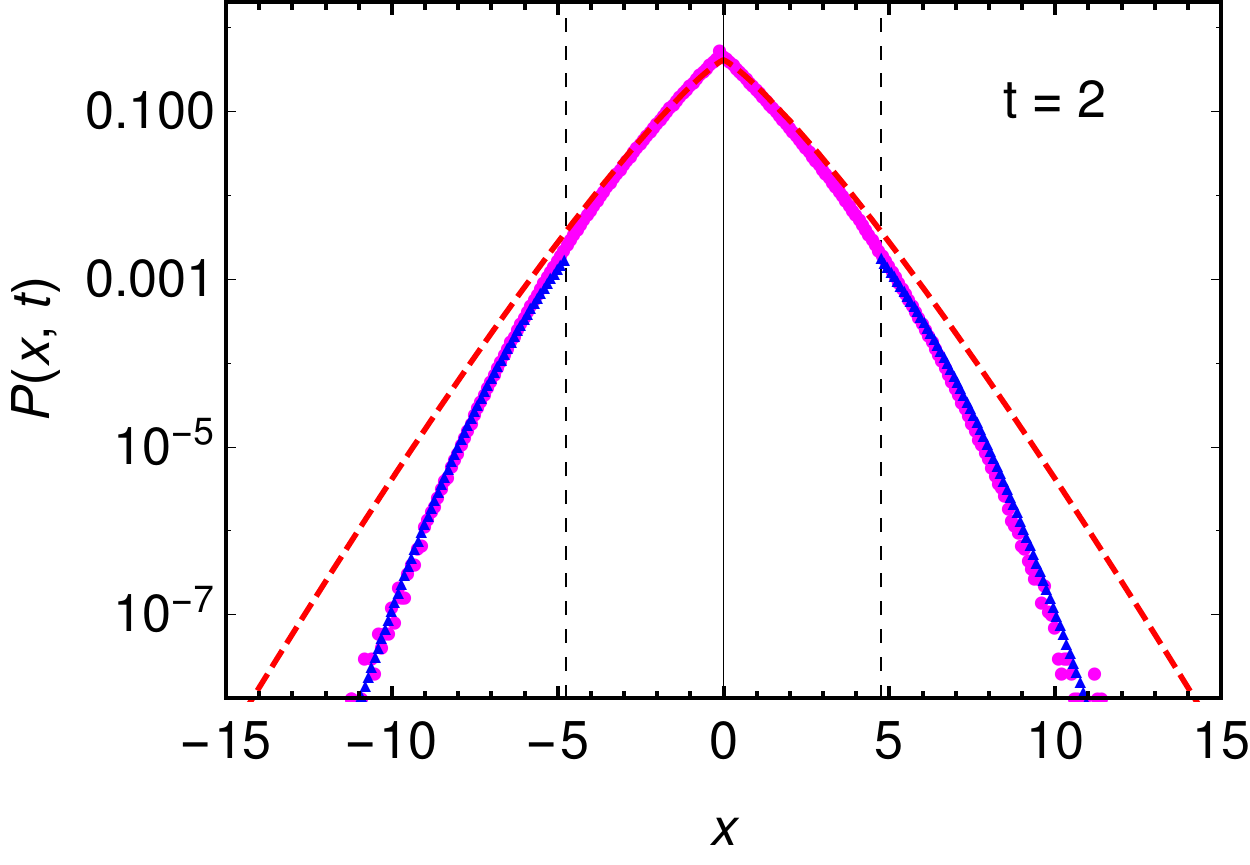}
\caption{(Color online) Dynamical relaxation
of a single Brownian particle under the time-dependent stochastic resetting to the fixed point at the origin is
shown. The resetting rate has the time dependence, according to \eref{rt-form}, with $\theta=0.5$ fixed for all the dynamical processes observed at different time scales $t=~1.0,~1.5,~2.0$ respectively. 
The intrinsic parameters for the system are : $~D = 1,~b_0=1$. The distribution is characterized by two regimes: inside one where the non-equilibrium steady state has already set in and the outside 
one where the system is still in a transient state. The front given by \eref{front} separates these two regimes and have been depicted by the dashed vertical lines in the figures. 
The simulation results, indicated by magenta circles, agree well with the theoretical results obtained from \eref{dyna_tr_LD}, for both the steady state (dashed red curves) and 
the transient (the blue triangles) regimes.}
\label{figr2}
\end{figure*}

The dominant asymptotic  large time, and large $|x-x_0|$  behavior is therefore given by the following large deviation form: 
\begin{eqnarray}
\mathbb{P}_r(x,t|x_0,0) &\sim&  \exp \left[{-t^{1+\theta}\mathbb{I}(u)} \right],~~\text{with}~~u=\frac{|x-x_0|}{t^{1+\theta/2}}, \nonumber \\
\text{and},~~ \mathbb{I}(u)&=&
 \begin{cases}
  & b_0 \left( \frac{\theta+2}{\theta+1}\right)\left( \frac{u^2}{4 D b_0}\right)^\frac{1+\theta}{2+\theta}~~\text{for}~~u<u^* \\
   & \\
  & \frac{b_0}{1+\theta}  + \frac{u^2}{4D}~~~~~~~~~~~~~~~~~~~~\text{for}~~u>u^*
 \end{cases} \label{dyna_tr_LD}
\end{eqnarray}
where $u^*=\sqrt{4 D b_0}$. The interpretation of this result is that
the inner region $u<u^*$ or equivalently, $|x-x_0| < \sqrt{4 D b_0}\,
t^{1+\theta/2}$ has a time-independent behavior and therefore already
has reached the non-equilibrium steady state form. This is the relaxed
regime.  On the other hand, the outer region $u>u^*$ or equivalently,
$|x-x_0| > \sqrt{4 D b_0}\, t^{1+\theta/2}$ is yet to equilibrate and
still in a transient state. Thus an equilibration front moves through
the system: the motion is superlinear if $\theta >0$; linear if
$\theta =0$ with speed $u^*=\sqrt{4 D b_0}$, and sublinear but
superdiffusive if $-1 < \theta <0$.  The case $\theta=0$ recovers the
results of \cite{MSS15}.

In Fig. \ref{figr2}, we observe this dynamical transition from
numerical simulations, where we plot the  distribution
$\mathbb{P}(x,t|x_0,t)$ as a function of $x$ for a given value of
$\theta$ at three different observation time scales.  For comparison
we plot the theoretical large deviation form of the distribution given
by~\eref{dyna_tr_LD} along with the simulation results. We find  very
good agreement between them in both the transient and the steady state
regimes.


\section{Survival Probability under time-dependent resetting}
\label{survival_probability}
We now look into the first passage probabilities of a diffusing
particle under the influence of  time-dependent stochastic
resetting. First passage properties are generically of importance
since they characterise the performance of search processes in various
contexts. The most intuitive and important observable is the first
passage time (FPT), for the particle to reach the origin say, which is
itself a stochastic quantity and the PDF of FPT has its own rich
characteristics.  In particular, one of the main goals is to optimize
the mean first passage time (MFPT) with respect to the system
parameters.  A review of the topic can be found in \cite{BMS13}.

The problem of computing the survival probability in presence of a
static target has been well studied for free Brownian motion (see
e.g. \cite{Majumdar:2005} for a review).
The MFPT was well investigated for diffusion under 
stochastic resetting at a constant rate  in \cite{EM11a,EM11b}.  It was found that the MFPT attains a
minima with respect to the rate in contrast to the case of a single
diffusive searcher in the absence of resetting.  Thus there in fact
exists an optimal resetting strategy which can be useful for all kinds
of persistence problems.  In this
section, we address the question whether this
strategy can be made even more efficient by introducing time-dependent
resetting.

To this end, we compute the survival probability in presence of
resetting at a time-dependent rate. To do so it is useful to introduce
$q(x_0,t)$ that defines the survival probability of a free Brownian
particle at $0$ until time $t$, starting from $x_0$. This result is
well known (see e.g. \cite{Redner}) and is given by
\bea
q(x_0,t)=\Erf\left[\frac{x_0}{\sqrt{4Dt}}\right].
\l{survival1}
\eea


Let $Q_r(x_0,x_r,t)$ be the probability that the Brownian particle
starting from $x_0$ survives until time $t$ without being
absorbed  at $0$ in the presence of stochastic resetting to
$x_r$. The  suffix `$r$' indicates the presence of resetting, as before. 
Without any loss of generality, once again we will assume $x_r=x_0$.

Using the \textit{first renewal} formalism, we now 
write the survival probability of the Brownian particle subject 
to time-dependent resetting
\bea
Q_r(x_0,x_0,t)&=&e^{-R(t)}~q(x_0,t)  \l{reset-time-survivalf} \nn \\
&+&\int_0^t d\tau_f~r(\tau_f) e^{-R(\tau_f)}q(x_0,\tau_f)~Q_r(x_0,x_0,t-\tau_f)~.
\eea
The first term on the RHS represents trajectories  in which
there has been neither reset to $x_0$ nor absorption at the
origin until time $t$. 
The integral in the   second term represents a sum over first reset
times $\tau_f$  and implies that there
has been no reset and no absorption  until time $\tau_f$ then 
 a reset between $\tau_f$ and $\tau_f+d\tau_f$.
The factor $r(\tau_f)~e^{-R(\tau_f)}~q(x_0,\tau_f)$
inside the integral  gives the probability
of no  resetting and no absorption up to time $\tau_f$ and 
a resetting event
between $\tau_f$ and $\tau_f+d\tau_f$.
This is multiplied with the  term
$Q_r(x_0,x_0,t-\tau_f)$ which simply implies that
no absorption occurs from $\tau_f$ to $t$.

Equation (\ref{reset-time-survivalf}) belongs to a Wiener-Hopf class of integrals
which can be solved in the Laplace space
\bea
\tilde{Q}_r(x_0,x_0,s)=\f{\tilde{q}_r(x_0,s)}{s\tilde{q}_r(x_0,s)-\tilde{k}_r(x_0,s)} ~,
\l{reset-time-survivalfLT}
\eea
where we have defined the following quantities
\bea
\tilde{q}_r(x_0,s)=\int_0^\infty dt~e^{-st}e^{-R(t)}~q(x_0,t) \label{Q_tilda}~, \\
\tilde{k}_r(x_0,s)=\int_0^\infty dt~e^{-st}e^{-R(t)}~\f{d}{dt}[q(x_0,t)]~.
\label{k_tilda}
\eea
The mean first passage time to the origin starting from $x_0$ in the presence of stochastic resetting to
$x_r=x_0$ is then given by
\bea
T(x_0)&=&-\int_0^\infty~dt~t~\f{\partial{Q_r(x_0,x_0,t)}}{\partial{t}} =\tilde{Q}_r(x_0,x_0,s \to 0),~~~
\l{MFPT1}
\eea
which upon using ~(\ref{reset-time-survivalfLT}) yields
\bea
T(x_0)&=&-\f{\tilde{q}_r(x_0,0)}{\tilde{k}_r(x_0,0)}. \l{MFPT3} 
\eea
Inserting the expressions of $\tilde{q}_r(x_0,0)$ and $\tilde{k}_r(x_0,0)$ from 
eqs. (\ref{Q_tilda}) and (\ref{k_tilda}) and performing some simple manipulations, 
one can write $T(x_0)$ in a compact form :
\bea
T(x_0)=- 4~I(\beta)~\left(\frac{d^2 I}{d \beta^2}\right)^{-1}~,~\text{with}~~\beta=\frac{x_0}{\sqrt{4D}}~.
\l{MFPTtime2}
\eea
The integral $I(\beta)$ is given by
\bea
I(\beta)=\int_0^\infty dt~{\rm e}^{-R(t)}~\Erf\left(\frac{\beta}{\sqrt{t}}\right). \label{I_beta}
\eea
For \textit{time-independent}  resetting rate $r(t)=r_c$, one can easily compute $I(\beta)$ to find 
$I(\beta)= \frac{1}{r_c}(1-e^{-\alpha_0x_0})$ where $\alpha_0=\sqrt{\f{r_c}{D}}$, is the
inverse distance diffused between two resetting
events. Now using this expression for $I(\beta)$ in ~(\ref{MFPTtime2}) we recover the result
\bea
T(x_0)=\f{1}{r_c}\left[ {\rm e}^{\alpha_0 x_0}-1 \right]~,
\l{MFPT4}
\eea
obtained in \cite{EM11a}. In \cite{EM11a}, it was also shown that there is an optimal choice of the 
constant value of the rate $r^*_c= (2.53964...)(D/x_0^2)$ for given initial/resetting position $x_0$ 
and diffusion constant $D$, such that the value of MFPT becomes minimum : $T^*(x_0)= (1.54414...)(x_0^2/D)$.

The question now is whether one can make the search process more efficient
by introducing time-dependent resetting rates. To answer this
question, we compute the MFPT $T(x_0)$ 
for two choices of rate functions in the following. Then varying the respective parameters in the two 
rate functions we find that in both the cases one can get smaller $T(x_0)$ than the minimum
$T^*(x_0)$ possible using a constant resetting rate $r_c$. 

\subsection{$r(t)=b_0 t$}
For this case, $R(t)=b_0t^2/2$ grows quadratically with time. Inserting this explicit form of $R(t)$ in 
~(\ref{I_beta}) and performing some simple variable changes one finds that 
\bea
I(\beta)= \sqrt{\f{2}{b_0}} J\left(\beta \sqrt[4]{\f{b_0}{2}} \right),
\eea
where $J(\ell)$ is defined in ~(\ref{J(l)}) and given explicitly in
Appendix \ref{sec:app}. Using this $I(\beta)$ in 
~(\ref{MFPTtime2}) one can find $T(x_0)$ for any given $b_0$, $x_0$ and $D$.
In \fref{figMFPT1}a, we plot this MFPT as a function of $b_0$ for $x_0=1$ and $D=1$. 
We observe a wide range of $b_0$ values for which  $T(x_0)$ is much smaller 
than the minimum possible MFPT \emph{i.e.} $1.54414...$ using a constant rate.


\begin{figure*}
\includegraphics[width=.3\hsize]{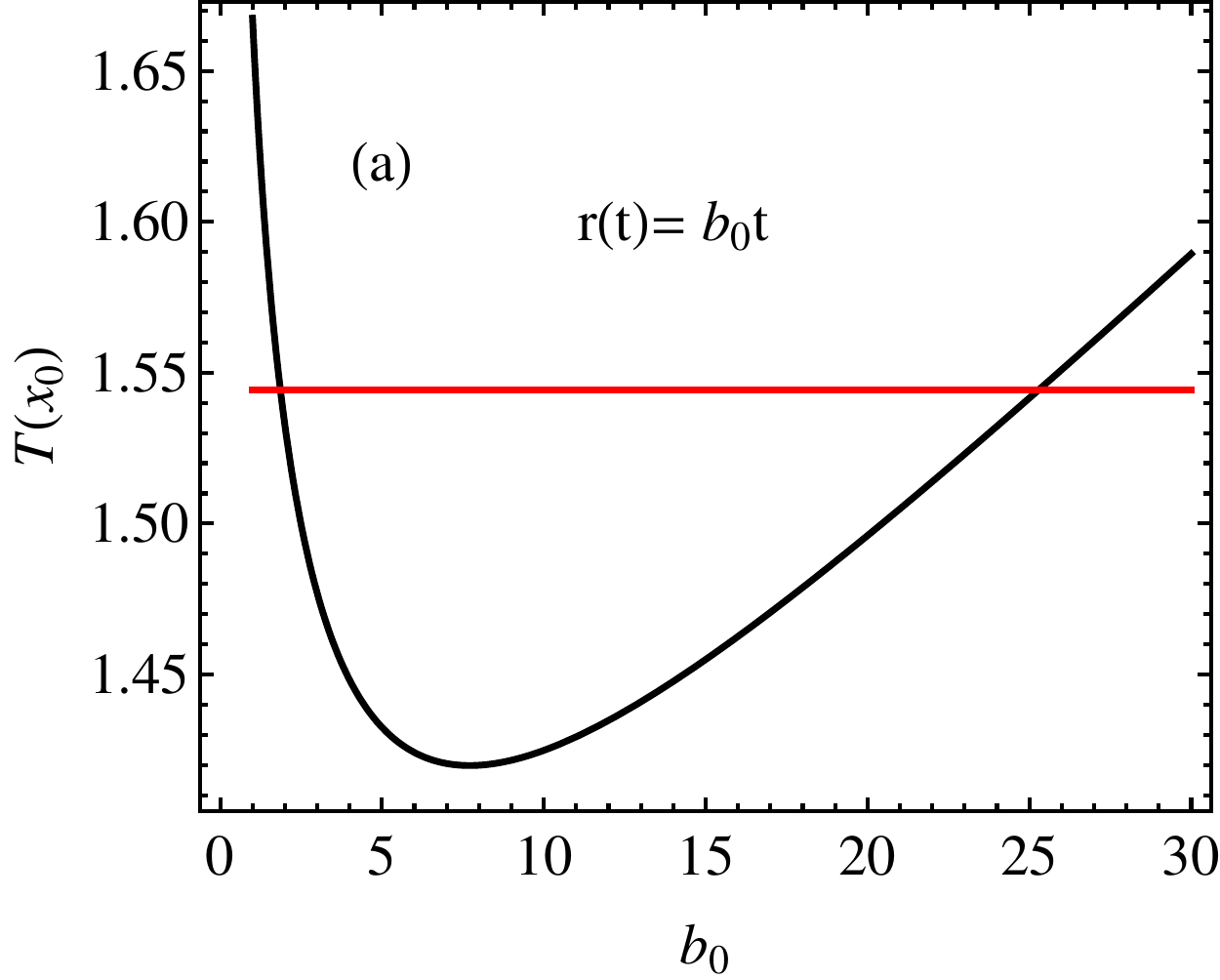}
\includegraphics[width=.3\hsize]{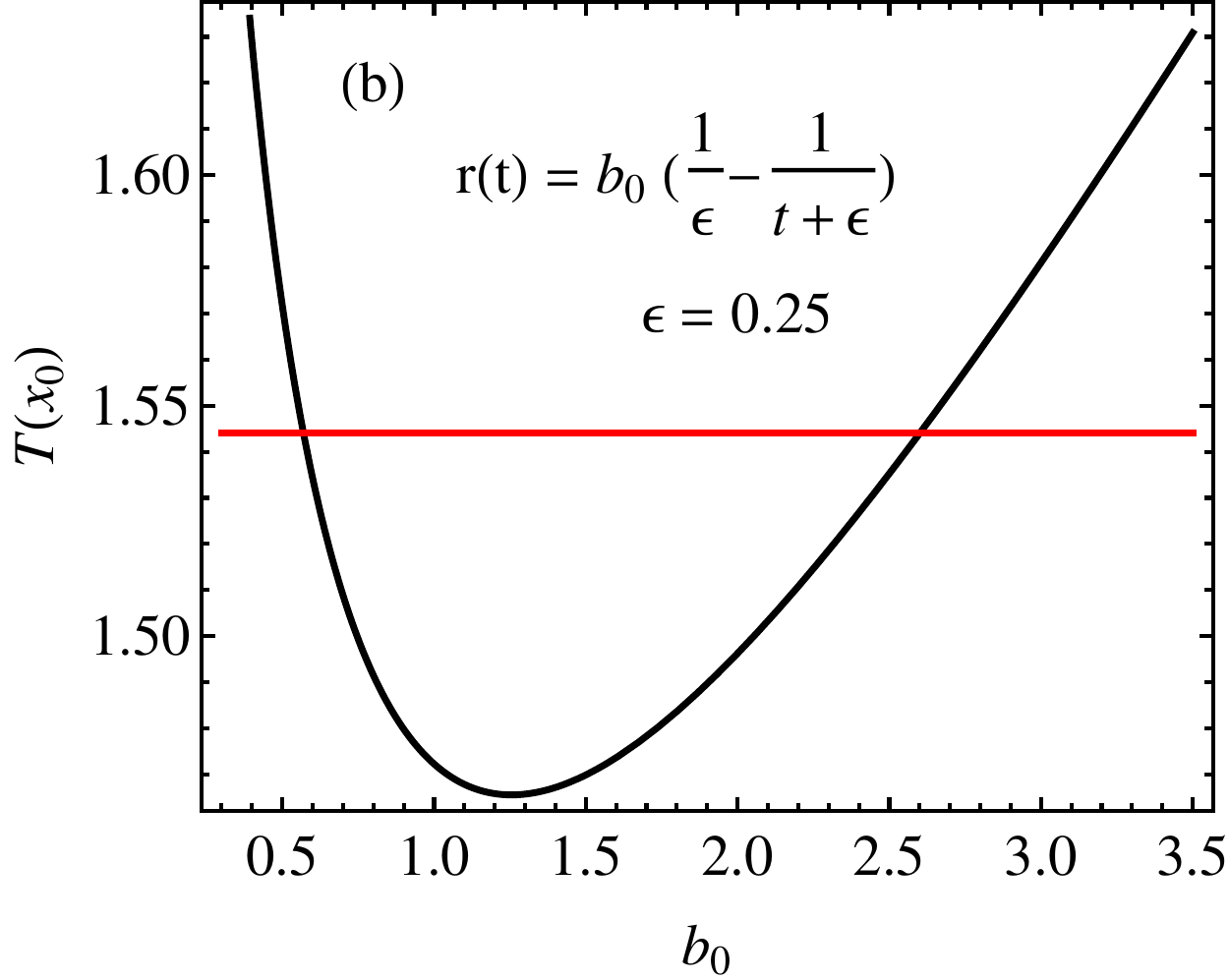}
\caption{(Color online) Plot of MFPT as function of the parameter $b_0$ for the following two reset rate functions : 
$r(t)=b_0~t$ and $r(t)=b_0~\left(1/\epsilon - 1/(t+\epsilon)\right)$. The parameter $b_0$ in each cases of $r(t)$ is adopted in appropriate units. 
The diffusion constant $D=1$ and the initial/resetting position has been fixed to $x_r=x_0=1$. We observe in both cases that there is a wide range 
of parameter for which the MFPT is much smaller than the minimum MFPT $T = 1.54414...$ 
(represented by the horizontal lines) possible using constant rate function $r(t)=r_c$ with the optimal value at $r_c^*$.}
\l{figMFPT1}
\end{figure*}


\subsection{$r(t)=b_0(\f{1}{\epsilon}-\f{1}{t+\epsilon})$}
In this case $R(t)$ is given by 
\bea
R(t)=\int_0^t d\tau~r(\tau)=b_0\bigg[ \f{t}{\epsilon}-\ln|1+\f{t}{\epsilon}| \bigg],
\l{ratetime2}
\eea
which, on inserting into  (\ref{I_beta}) and performing some simple manipulations, yields
$I(\beta)=\epsilon~\mathcal{J}_{b_0}(\beta/\sqrt{\epsilon})$. The function $\mathcal{J}_{b_0}(\ell)$ 
is given in ~(\ref{J_b_0(l)}). Using Eq. (\ref{MFPTtime2}), one thus obtains  
\begin{equation}
 T(x_0)= \f{\sum_{k=0}^\infty \sum_{m=1}^\infty \f{(-1)^{k+m}}{k!m!}\left(\f{2x_0}{\sqrt{4D\epsilon}}\right)^m(b_0)_k\left(\f{m-2}{2}\right)_k b_0^{m/2-k-1}}
 {1-\sum_{k=0}^\infty \sum_{m=1}^\infty \f{(-1)^{k+m}}{k!m!}\left(\f{2x_0}{\sqrt{4D\epsilon}}\right)^m(b_0)_k\left(\f{m}{2}\right)_k b_0^{m/2-k}},
\end{equation}
where the symbol $(a)_k=a(a-1)(a-2)...(a-k+1)$. In \fref{figMFPT1}b, we plot this MFPT as a function of $b_0$ for fixed $\epsilon$, $x_0$ and $D=1$. 
Here also we observe a wide range of $b_0$ values for which the $T(x_0)$ is much smaller 
than the minimum possible MFPT \emph{i.e.} $1.54414...$ using constant rate.


\section{Optimal resetting rate function}
\label{optimal-r(t)}
In this section we address the question of optimizing the search
processes by minimizing the MFPT. We ask: What is the optimal form of the
time-dependent resetting rate $r(t)$?
Equation (\ref{MFPT3}) is the primary equation for the MFPT and
to seek the optimized form
let us take the
functional derivative in \eref{MFPT3} with respect to the function $r(t)$ at time $t'$
\begin{eqnarray}
\frac{\delta T(x_0)}{\delta r(t')}&=&\frac{ \int_{t'}^\infty dt~ {\rm
    e}^{-R(t)}q(x_0,t)}{\tilde k_r(x_0,0)} \nonumber \\ 
    &&- \frac{ \tilde q_r (x_0,0)}{ \left[\tilde k_r (x_0,0)\right]^2}
\int_{t'}^\infty dt ~{\rm
    e}^{-R(t)}\frac{\partial q(x_0,t)}{\partial t}~.
\label{fd}
\end{eqnarray}
Setting the RHS to zero yields an equation which must be satisfied for
all $t'$ in order to have an extremal resetting function. However, it
is clear that according to (\ref{fd}) one cannot find such a function:
if the RHS of (\ref{fd}) is to be a constant value zero for all $t'$,
then its derivative with respect to $t'$ must vanish which yields
\begin{equation}
 {\rm e}^{-R(t')}\left[\tilde k_r (x_0,0)
 q(x_0,t')
-
 \tilde q_r (x_0,0)
 \left.\frac{\partial q(x_0,t)}{\partial t}\right|_{t=t'} \right] =0~,
\end{equation}
or, rearranging,
\begin{equation}
\frac{ 1 }{q(x_0,t')}\left.\frac{\partial q(x_0,t)}{\partial t}\right|_{t=t'} 
=  \frac{
 \tilde k_r (x_0,0)}
{ \tilde q_r (x_0,0)}~,
\end{equation}
and this equation cannot be satisfied for all  $t'$ since the RHS is
constant
but the LHS is a  function of $t'$. Indeed the LHS does not depend on $r(t)$.

Since the optimal resetting function does not extremise the MFPT
everywhere (i.e. it does not set the functional derivative to zero everywhere), it
must instead  involve values of rates at the boundary of the
allowed rate space i.e. it must involve $r=0$ (since a rate cannot
be negative) and $r \to \infty$.  This leads us to conjecture that a
possible form for the
optimal resetting function is 
\begin{eqnarray}
\label{optr1}
r(t) =
\begin{cases}
 &0\quad\mbox{for}\quad t< t^*\\
 &\infty \quad\mbox{for}\quad t> t^*.
\end{cases}
\end{eqnarray}
That is, resetting does not occur up until time $t^*$ then occurs
instantaneously.
Or in other words there is  deterministic resetting with period $t^*$.

For $r(t)$ of the form (\ref{optr1}) the MFPT is given by
\begin{equation}
T(x_0)= \frac{\int_0^{t^*} dt~ q(x_0,t)}{1-q(x_0,t^*)}~.
\label{T*}
\end{equation}
Extremising this expression with respect to $t^*$
yields the condition
\begin{equation}
\frac{\int_0^{t^*} dt' ~q(x_0,t')}{1-q(x_0,t^*)}= -\frac{q(x_0,t^*)}{\left. \frac{\partial q(x_0,t)}{\partial t}\right|_{t=t^*}}~,
\label{t*}
\end{equation}
where, as usual, $q(x_0,t)=\text{erf}\left(x_0/\sqrt{4 D t}\right)$. Note that the above equation can be expressed solely in terms of the 
rescaled variable $z=t/\beta^2$ as 
\begin{equation}
\frac{\int_0^{z^*} dz' ~\text{erf}\left({1}/{\sqrt{z'}}\right)}{1-\text{erf}\left({1}/{\sqrt{z^*}}\right)}
= -\frac{\text{erf}\left({1}/{\sqrt{z^*}}\right)}{\frac{\partial \text{erf}\left({1}/{\sqrt{z^*}}\right)}{\partial z^*}}~,
\label{z*}
\end{equation}
where $\beta=x_0/\sqrt{4D}$ as before. This implies that $t^*=z^*~\frac{x_0^2}{4D}$ where $z^*$ is the solution of the above equation.
One can solve this equation numerically to find $z^*=1.834011077...$
Using $t^*=z^*~\frac{x_0^2}{4D}$ in (\ref{T*}) one finds that the optimum MFPT is given by $T(x_0)= 5.34354...({x_0^2}/{4D})$ which 
is much smaller than the minimum MFPT $6.17655...({x_0^2}/{4D})$ possible for any given $x_0$ and $D$ using constant reset rate 
(see fig. \ref{figMFPT2}). Secondly, the MFPT $T(x_0)$ and $t^*$ constitute a linear relation $T(x_0)= 2.91358...t^*$,
for any given $x_0$ and $D$ (see fig. \ref{figMFPT2}).

\begin{figure*}
\includegraphics[scale=.255]{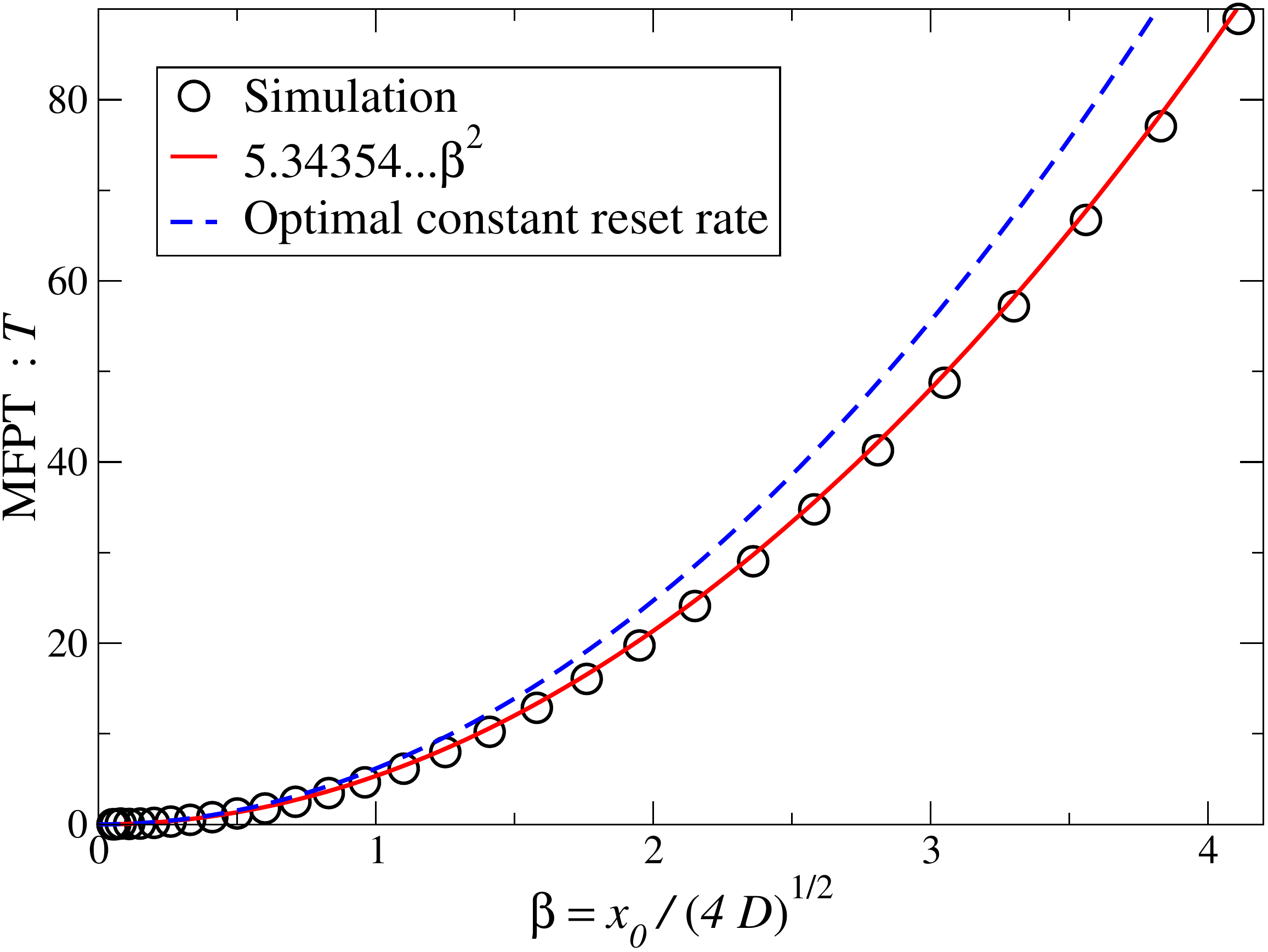}
\includegraphics[scale=.3]{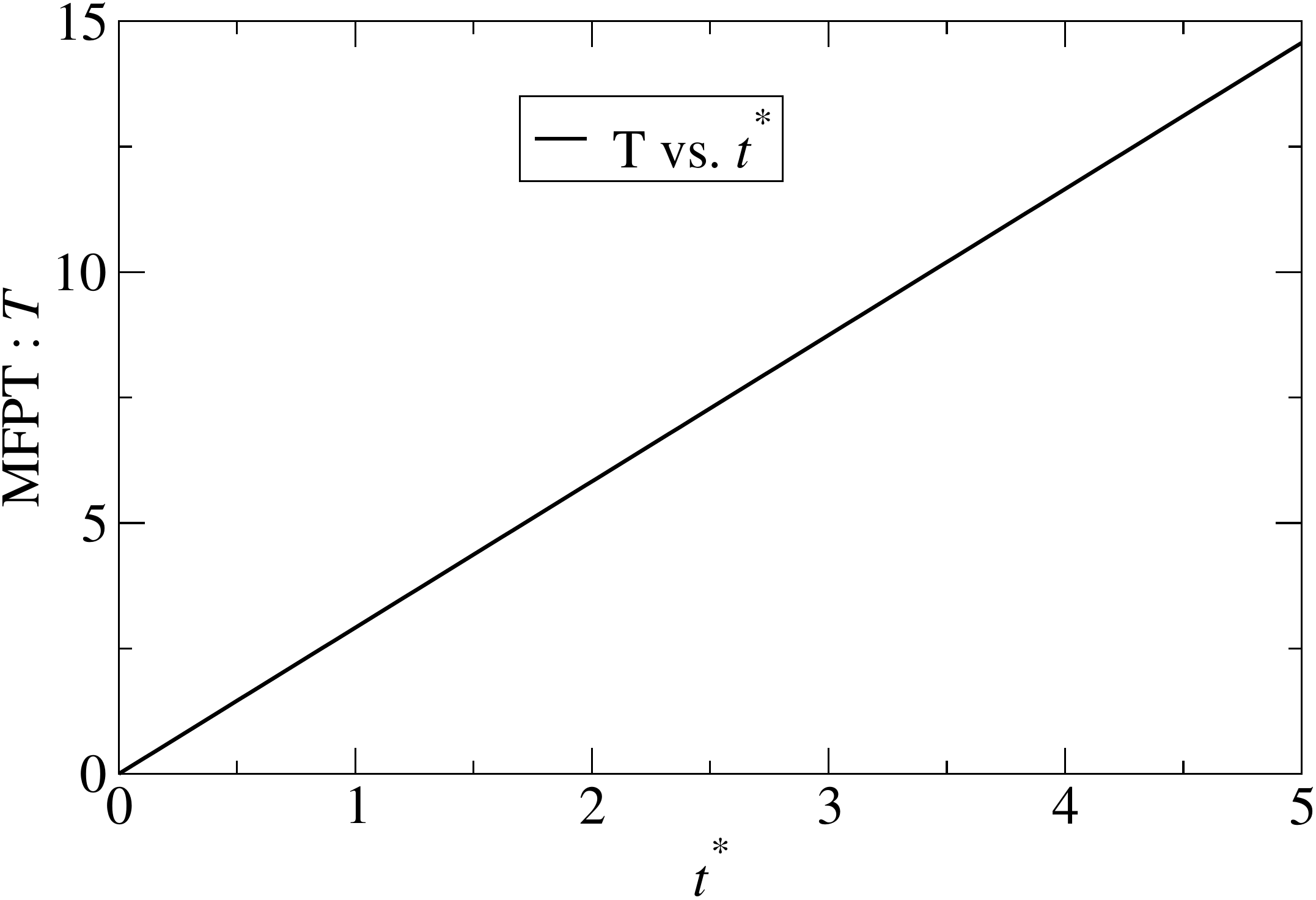}
\caption{(Color online) Left Panel: We plot MFPT $T(x_0)$ as a function of $\beta=x_0/\sqrt{4D}$ for the optimal resetting rate function \eref{optr1}. The circles are 
obtained from numerical simulation whereas the red solid line represents the theoretical expression $T=5.34354...\beta^2$. The blue dashed line corresponds to the same 
for constant resetting rate with optimal value $r(t)=r_c^*=2.53964.../\beta^2.$ Right Panel: The linear relation between the MFPT $T(x_0)$ and $t^*$, as mentioned in the text, has been plotted.}
\l{figMFPT2}
\end{figure*}

Furthermore we can check that
the   rate  function $r(t)$ given by  \eref{optr1}
with $t^*=z^*~\frac{x_0^2}{4D}$, locally optimizes the MFPT with
respect to variations of the rate function.
First note that
the functional derivative of the MFPT $\to 0$ for $t'>t^*$.
For $t'<t^*$, after some simplification, we obtain 
\begin{equation}
\frac{\delta T(x_0)}{\delta r(t')} =
\frac{1}{[1- q(x_0,t^*)]^2} ~K(x_0,t') ~,
\end{equation}
where
\begin{equation}
K(x_0,t')=
(1- q(t^*)) \int_0^{t'} dt ~q(x_0,t)
- (1- q(t')) \int_0^{t^*} dt~ q(x_0,t)~. \label{Kdef}
\end{equation}
Now consider  the function $K(x_0,t')$: it takes value 0  and has positive derivative
at $t'=0$ while
at $t'=t^*$ it takes value zero and has derivative equal to zero. In the domain
$ 0< t' < t^*$, the function  $K(x_0,t')$ has  a single turning point as can be checked
numerically.
Hence, in the domain $0< t'<t^*$, the function $K(x_0,t') >0$,
is strictly positive 
as is the functional derivative $\displaystyle \frac{\delta T(x_0)}{\delta r(t')} >0$. Thus
$r(t)$ given by  \eref{optr1}
is evidently a locally optimal rate function with respect to the MFPT.


\section{Conclusions}
\noindent
In this work we have considered a simple stochastic
system of a Brownian particle subjected to a time-dependent resetting rate to a fixed
position $x_r$. This is in contrast with  earlier studies
where a time independent rate was chosen.
The rate of resetting $r(t)$ is a function of the time $t$ since the
last reset event.
We have seen that for steady states to exist, the function $r(t)$ must respect
certain features such as
either increasing with time, tending to some finite constant as $t\to
\infty$ or decay to zero sufficiently slow.
A sufficient  condition for the steady state is given by (\ref{sscond}).
In this context, various cases for the steady state distributions have been worked out in
Section \ref{SS-time-dep-r}.

We have also considered the late time relaxation behavior of the
probability distribution for the specific choices of rate function
$ r(t) \propto t^\theta$ with $\theta > -1$. In analogy to recent work
\cite{MSS15} we have shown that an inner core region around the resetting point settles into
the steady state distribution and  the boundaries of the inner core region
is still in transient state and
move outwards towards the tails of the distribution with time.
Thus an `equilibration' front moves through the
system and its motion is superlinear if $\theta >0$;  linear if
$\theta =0$ (recovering the results of \cite{MSS15})
and sublinear but superdiffusive if $-1 < \theta <0$.

Finally we have considered persistence properties of the Brownian
motion in the presence of the time dependent resetting process. In
particular, we have studied extensively the first passage time
properties of the Brownian particle to reach the origin.  We have
derived a formula (\ref{MFPTtime2}) for the mean first passage time.
A primary motivation has been to optimize the mean first passage time
with respect to the rate functions. To this end, we have studied
thoroughly the optimal properties of the mean first passage time and
have shown that a threshold function (\ref{optr1}) with an optimised
threshold $t^*$ is at least locally optimal for the problem of
minimizing the mean first passage time.  It would be of great interest
to see whether this is also the globally optimal function and if so,
to provide a rigorous proof of this.

Other interesting generalizations would be to consider, for example,
oscillatory reset functions or else random resetting rates chosen from
a specified distribution.  It would also be interesting to look at the
properties of the entropy generation in the system subject to a
resetting mechanism. Another propitious future direction might be to
study the effects of time-dependent resetting on the dynamics of
interacting multi-particle systems, which could prove useful in the
context of cellular biology.

\begin{acknowledgements}
\noindent
The authors thank the Galileo Galilei Institute for Theoretical Physics Program 
`Advances in Nonequilibrium Statistical Mechanics' under-which this work
was started. MRE is partially supported by EPSRC grant 
EP/J007404/1 and thanks the Weizmann Institute for the award of a
Weston Visiting Professorship during which the work was finished.
\end{acknowledgements}


\section{Appendix}
\noindent
\label{sec:app}
It has been shown in the main text that the steady state formula
for the linear rate process can be written in terms of the
function  $J(\ell)$ as mentioned in \eref{J(l)}. 
The integral $J(\ell)$ can be computed exactly using Mathematica and it is expressed in terms of hypergeometric functions as 
\begin{eqnarray}
 J(\ell) &=& \frac{\ell ~\Gamma \left(\frac{1}{4}\right) \,
   _1F_3\left(\frac{1}{4};\frac{1}{2},\frac{3}{4},\frac{5}{4};-\frac
   {\ell^4}{4}\right)}{\sqrt{\pi }}+\frac{4 \ell^3 ~\Gamma
   \left(\frac{3}{4}\right) \,
   _1F_3\left(\frac{3}{4};\frac{5}{4},\frac{3}{2},\frac{7}{4};-\frac
   {\ell^4}{4}\right)}{3 \sqrt{\pi }} \nonumber \\ 
   &&-2 \ell^2 \,
   _1F_3\left(\frac{1}{2};\frac{3}{4},\frac{5}{4},\frac{3}{2};-\frac
   {\ell^4}{4}\right)~.
\end{eqnarray}

\end{document}